\documentclass[preprints,article,accept,moreauthors,pdftex]{Definitions/mdpi} 

\usepackage{bm}
\usepackage{bbm}
\usepackage{listings}
\usepackage{subfig}
\usepackage{amsmath}
\usepackage{amssymb}
\usepackage{amsthm}
\usepackage{booktabs}
\usepackage{hyperref}
\usepackage{erewhon} % serif
\usepackage[sf,scale=1]{libertine}
\usepackage{fourier}
\usepackage[T1]{fontenc}
\usepackage{algorithm}
\usepackage{algorithmic}
\hypersetup{colorlinks,citecolor=teal}

\firstpage{1} 
\pubvolume{xx}
\issuenum{1}
\articlenumber{1}
\pubyear{2022}
\copyrightyear{2022}
\history{}

\newcommand*\diff{\mathop{}\!\kern0pt\mathrm{d}}

\Title{Roughness of the Implied Volatility}
\Author{Fabien {Le Floc'h}}
\AuthorNames{Fabien {Le Floc'h}}
\address{}
\corres{Correspondence: fabien@2ipi.com}
\abstract{The measures of roughness of the volatility in the litterature are based on the realized volatility of high frequency data. Some authors show that this leads to a biased estimate, and does not necessarily indicate roughness of the underlying volatility process. Here, we attempt to measure the roughness of the implied volatility of short term options, as well as of the VIX index, and evaluate whether they may be more appropriate proxies of the underlying instant volatility.}
\keyword{implied volatility, rough volatility, stochastic volatility}

\begin{document}

\section{Introduction and short review}
It is well-known that the assumption of constant volatility in the Black-Scholes model for pricing financial contracts is wrong and may lead to serious mispricing, especially for any exotic derivative contracts. A classic approach is to use a deterministic local volatility model \citep{dupire1994pricing} in order to take into account the variation both in the time dimension and in the underlying asset price dimension. But the model is still not realistic in terms of forward smile (the implied volatilities of forward starting options). A stochastic volatility component must be added to correct for it \citep{lipton2002volsmile}. More recently, the concept of rough volatility emerged in many academic papers. Instead of using a classic Brownian motion for the stochastic volatility process, a fractional Brownian motion is used. The idea of using a fractional Brownian motion for financial time-series can be traced back to Mandelbrot, but it is only relatively recently that it has been proposed for the volatility process (and not the stock process) \cite{gatheral2018volatility}.

After many published papers on the subject, \citet{cont2022rough} raise the question whether using a rough volatility process makes sense from a practical point of view. Indeed, they show that the Hurst index measured is different for the realized volatility (even based on very short sampling intervals) and the instantaneous volatility. The authors make the conjecture that is is all due to the discretization error (a.k.a. the microstructure noise), and that in fact, a standard Brownian motion is compatible with the low Hurst index of the realized volatility.

As mentioned in their conclusion, they are not alone in showing that the rough volatility assumption may not be justified, \citet{rogers2019things} has also published a note with the same conclusion.

\citet{fukasawa2019volatility} also attempt to answer the question: is volatility rough? They present the same problem with the estimation method of \citet{gatheral2018volatility}, namely that their measure may actually not measure the true Hurst index $H$, and this could be why it is always close to $H=0.1$, regardless of the market index. It is shown that for specific models, the measure introduced in \citep{gatheral2018volatility} is dependent on the period length considered (5 minutes vs. 1 minute lead to different Hurst indices). The authors therefore devise a more robust (and much more complex) way to estimate the Hurst index of time series, which is not so sensitive to the period length considered. They find that the market volatility, measured through the realized variance proxy is rough, with a Hust index $H=0.04$, is rougher than what Jim Gatheral suggested.

The study of \citet{gatheral2018volatility}, which motivated the use of rough volatility, was based on “biased” estimates of the Hurst index of the volatility of high frequency data as evidenced in the three papers cited above. Clearly then, this justification does not hold. \citet{fukasawa2019volatility} suggest the volatility is rougher, and it is slightly surprising that \citet{cont2022rough} do not take more those results into account. They shortly discuss it, but then go on proposing another estimate, one which seems to lead to the same results as Rosenbaum and Gatheral. To their credit, \cite{cont2022rough} show that the microstructure noise is far from IID, while \citet{fukasawa2019volatility} assume it is close to IID in their method. Perhaps their indirect conclusion is that \citet{fukasawa2019volatility} estimates are also biased (i.e. measuring something else - although the results presented in their paper seem solid).

The rough volatility models offer other interesting features such as the variation of the at-the-money skew in time, which mimicks the market implied skews in practice, a feature not easily achieved by stochastic volatility models (this typically requires a multidimensional stochastic volatility process to approach it crudely).

This short note firstly attempts to reproduce some of the results presented in \cite{cont2022rough}, and secondly applies their roughness estimator to other inputs, such as at-the-money implied volatilities as well as VIX index quotes.

After writing the first version of this note, we found out that \citet{livieri2018rough} also look at the Hurst index of short-term at-the-money implied volatilities, put in evidence and explain the upward bias of this latter proxy. We however do not find exactly the same results.

\section{Reproducing Cont et al. results}
The 5-minutes SPX realized volatility is freely available in CSV format at the \hyperlink{https://realized.oxford-man.ox.ac.uk/data}{Oxford-Man Institute of Quantitative Finance} and it is thus relatively straightforward to reproduce the numbers presented in \citep{cont2022rough}. The estimate of the Hurst index for a path $X$ observed on the time interval $[0,T]$ reads
\begin{equation}
	\hat{H}_{L,K} = \frac{1}{\hat{p}_{L,K}(X)}\,,\label{eqn:H_pestimate}
\end{equation}
where $\hat{p}_{L,K}$ is the solution of 
\begin{equation}
	W(L,K,p_{L,K}(X),T,X) = T\,,
\end{equation}
with 
\begin{equation}\label{eqn:pvariation}
	W(L,K,p,T,X) = \sum_{\pi^K \cap [0,T]} \frac{\left|X(t_{i+1}^K)-X(t_i^K)\right|^p}{\sum_{\pi^L \cap [t_i^K,t_i^{K+1}]} \left|X(t_{j+1}^L)-X(t_j^L)\right|^p} \left( t_{i+1}^K - t_i^K\right)\,,
\end{equation}
and $\pi^L$ is a partition of $[0,T]$ of $L$ elements, $\pi^K$ is a subpartition of $\pi^L$.

\begin{figure}[h!]
	\begin{center}
		\subfloat[][from different starting dates.]{\includegraphics[width=0.5\textwidth]{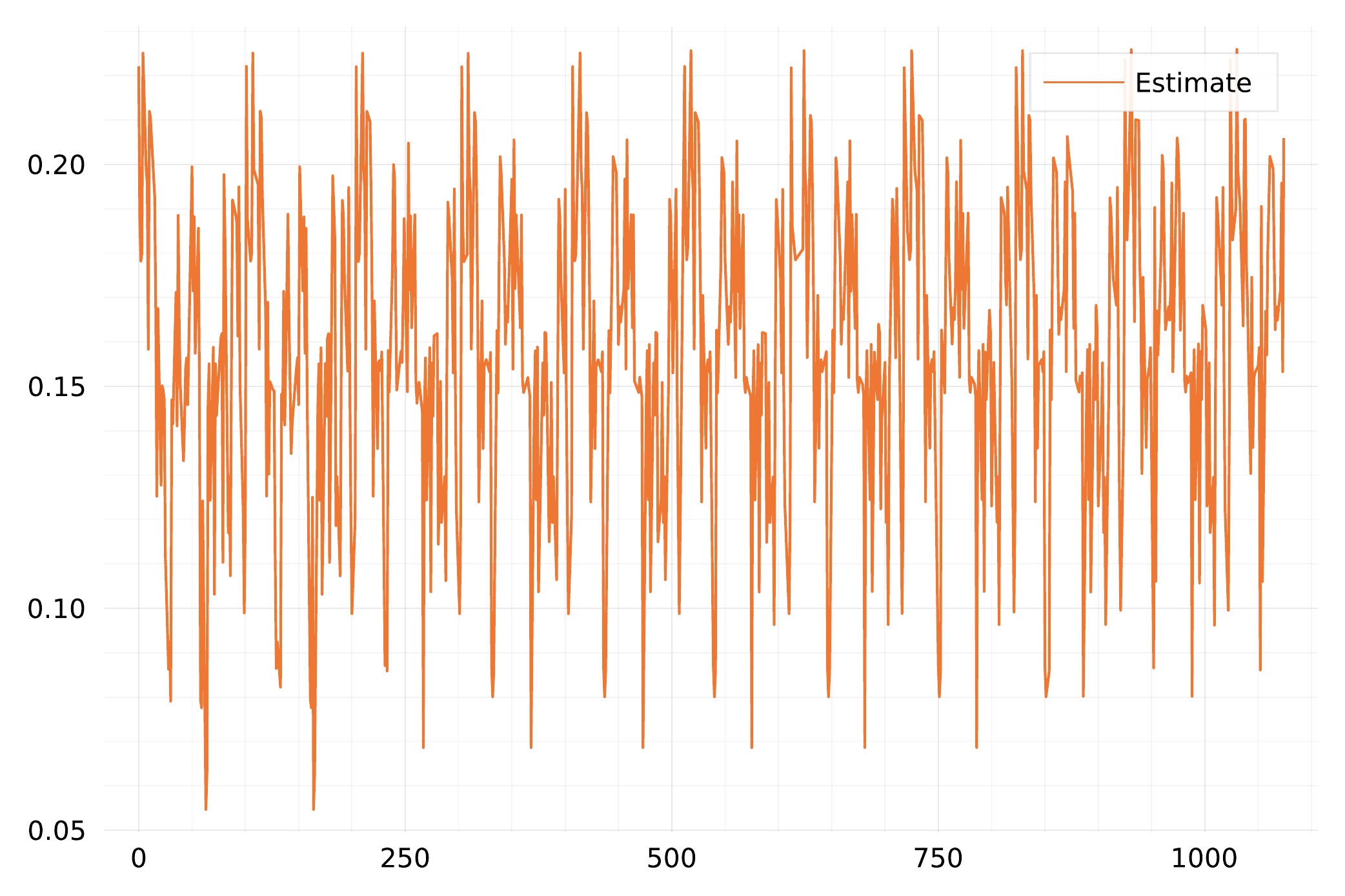}}
		\subfloat[][density. The mean is 0.158 with a standard deviation of 0.034.]{\includegraphics[width=0.5\textwidth]{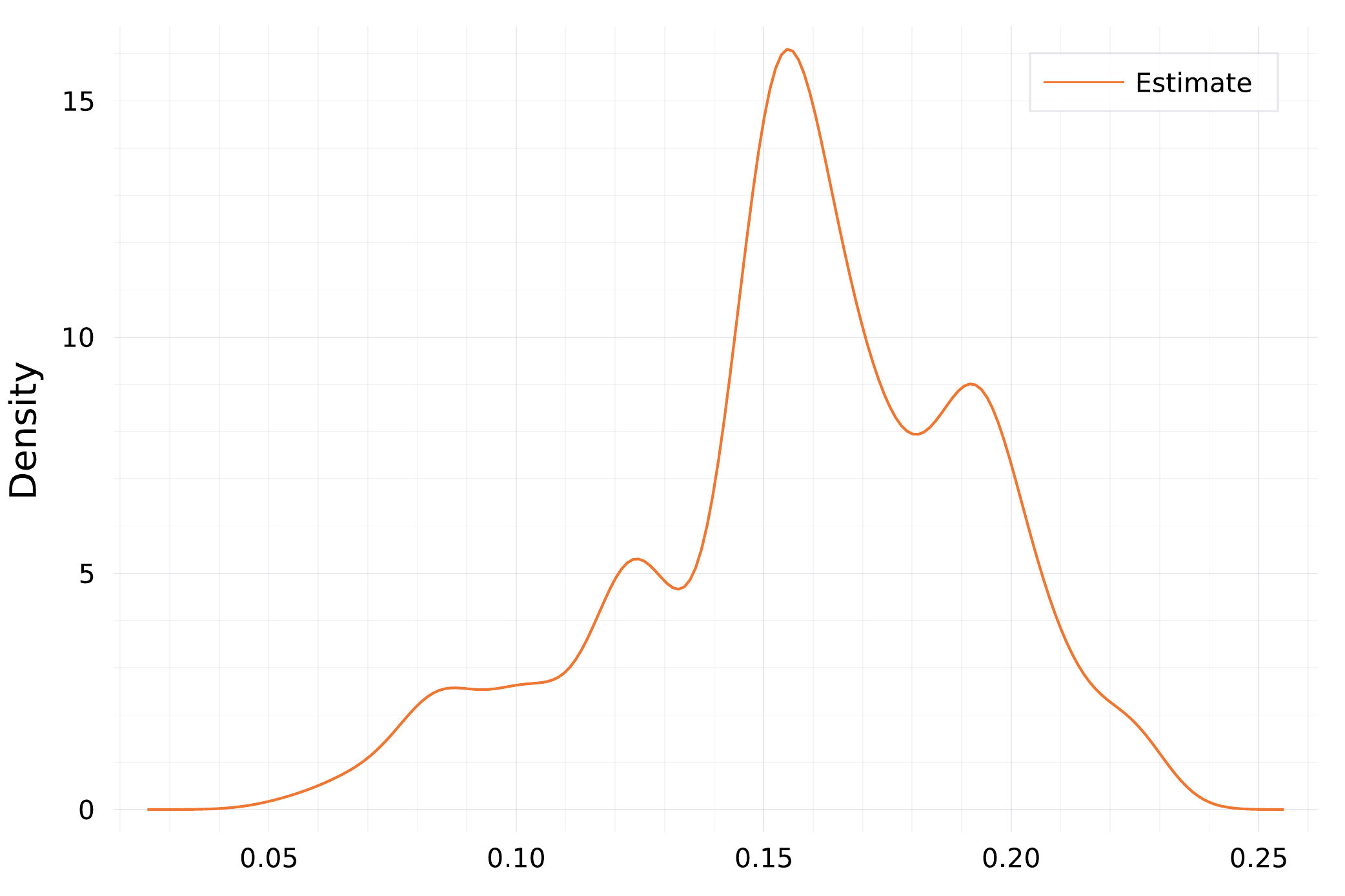}}
	\end{center}
	\caption{ SPX historical vol roughness index sliding estimate with $K=70$.\label{fig:spx_historical_vol_rough}}
\end{figure}

In practice we use a uniform partition of $L=K^2$ intervals of $[0, T]$. Equation (\ref{eqn:pvariation}) then reads
\begin{equation*}
	W(L,K,p,T,X) = \sum_{k=1}^K \frac{\left|X(t_{(k-1)K+K+1})-X(t_{(k-1)K+1})\right|^p}{\sum_{l=(k-1)K+1}^{(k-1)K+K} \left|X(t_{l+1})-X(t_l)\right|^p} \left( t_{(k-1)K+K+1} - t_{(k-1)K+1}\right)\,.
\end{equation*}
Using a sampling of $K=75$ and $L=75^2$, leads to a Hurst index $H=0.181$. The paper uses $K=70$ and $L = 70^2$, and \cite[Figure 19]{cont2022rough} states $H=0.187$. 
It turns out that there are more than $L$ observations in the time-series, and, with $K=70$, our measured roughness index is $H=0.222$, starting from the first observation (year 2000), up to the observation $L+1$. But it is possible to slide this window and compute the roughness index at each starting point. The results are enlightening (Figure \ref{fig:spx_historical_vol_rough}). 

The mean estimated roughness is 0.158 with a standard deviation of 0.034. The results of the paper are within one standard deviation. But it shows it may be dangerous to use a single window for the estimate. Figure \ref{fig:spx_historical_vol_rough} displays regular peaks, which are located $K$ elements apart. Indeed, moving from $K$ elements changes only one item of the main sum (the first sub-sum is replaced by a new sub-sum), while moving from one element changes all the items of the main sum.

\begin{figure}[h!]
	\begin{center}
		\subfloat[][from different starting dates.]{\includegraphics[width=0.5\textwidth]{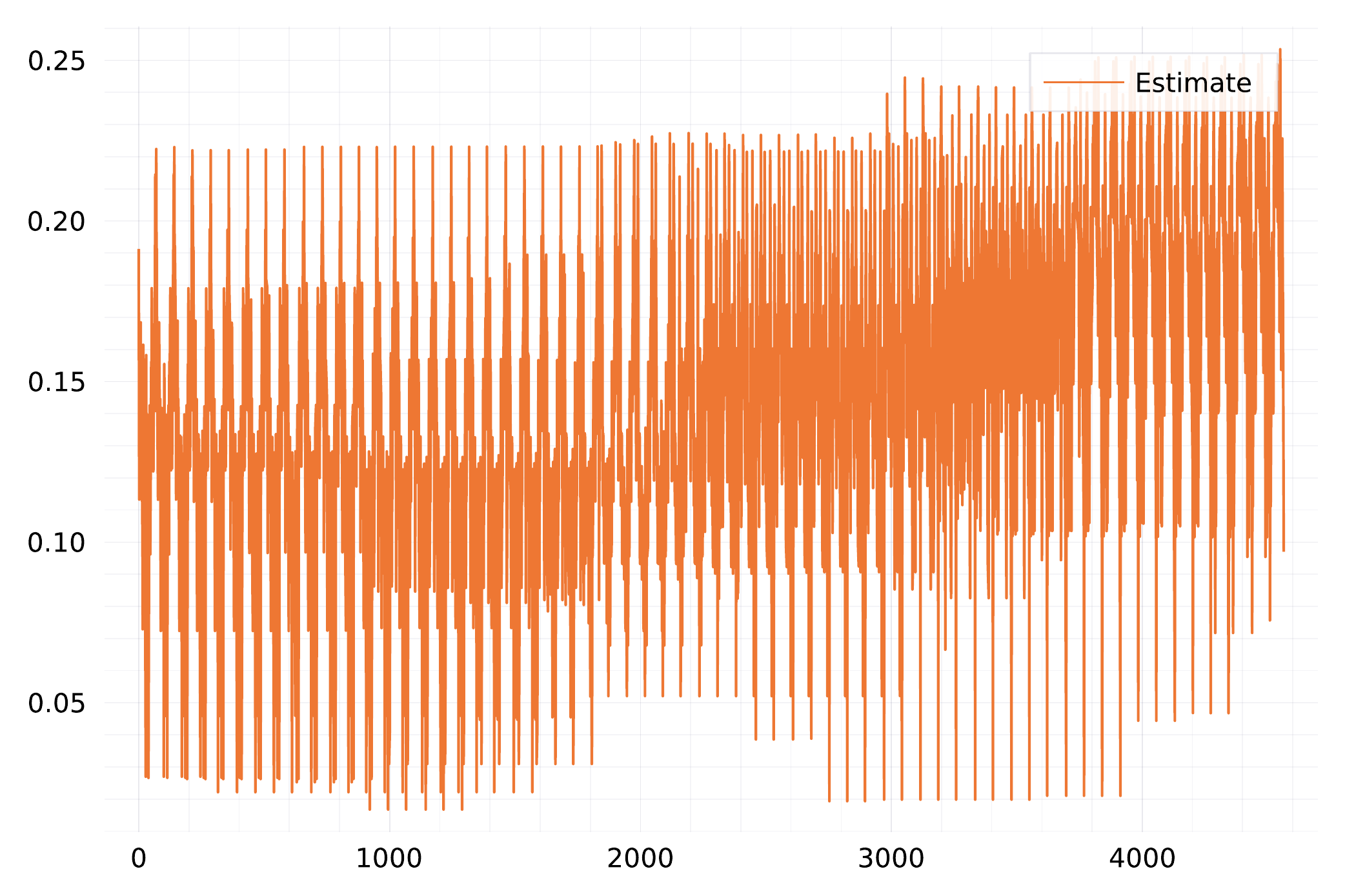}}
		\subfloat[][density. The mean is 0.145 with a standard deviation of 0.048.]{\includegraphics[width=0.5\textwidth]{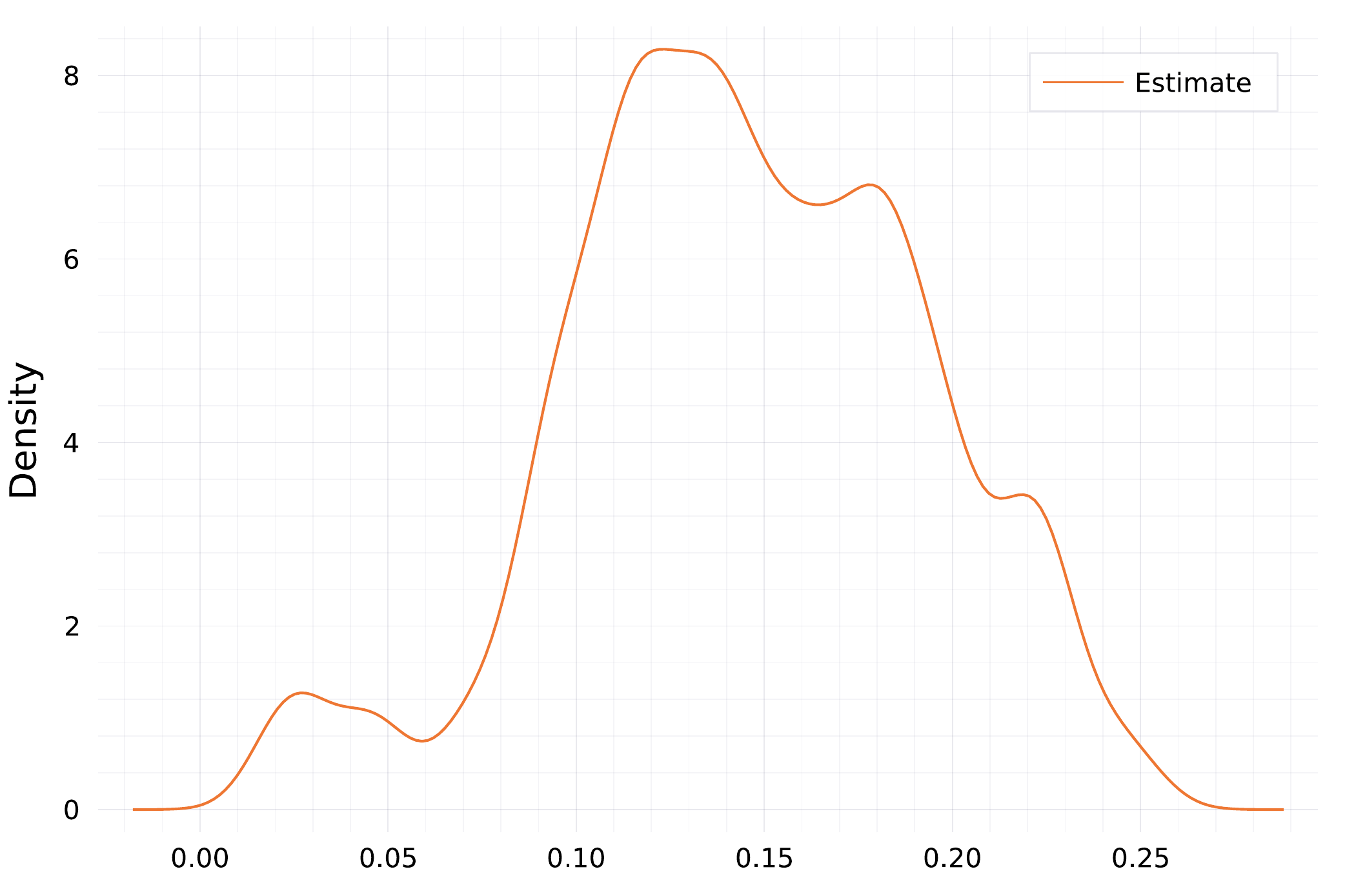}}
	\end{center}
	\caption{ SPX historical vol roughness index sliding estimate with $K=50$.\label{fig:spx_historical_vol_rough50}}
\end{figure}

If we use a smaller window size $K=50$, $L=50^2$, we can see regimes of roughness in time (Figure \ref{fig:spx_historical_vol_rough50}).

Finally, the measure on SPX closing prices resulted in an estimate $H=0.5$, for the two window sizes. The SPX price process follows the standard Brownian motion statistics.

\section{What about the Heston model?}
What is the estimate of the roughness index on a simulated stochastic variance path? Using one time-step per day for 5 years, and $K=25$, we find $H=0.505$ with a standard deviation of 0.032 for the roughness of the instantaneous variance, as well as for the one of the one-month at-the-money implied volatility (using $v(t)$ and constant Heston parameters in time). This is, as expected, the standard roughness of the underlying Brownian motion and there is no perceptible difference in the results between the implied volatility and the instantaneous volatility.

%There is actually a paper on option prices roughness equivalent to our.
%is it measuring noise of MC simulation??? since at every point, the iv is obtained from an MC sim. M=10K in paper. TODO try this out. The paper has actual analytical proofs as to why roughness is modified by IV maturity.

\section{Measure of roughness of the implied volatility}
\citet{cont2022rough} explain that using the 5 minutes realized volatility is not necessarily a good proxy to measure the true underlying instant volatility roughness. What happens if we use the daily quotes of the short term at-the-money (ATM) implied volatility as a proxy?

In a given stochastic volatility model, the short term ATM implied volatility is approximately the instant volatility of the stochastic volatility process. It turns out that there is some freely available data from the Nasdaq for the Microsoft stock (MSFT), but only for the period 2015-01-02 to 2018-12-32 (four years). This constitutes 1006 samples. For the 30-days implied volatility, with $K=28$, the data implies a roughness $H=0.448$, close to 0.5. 

\begin{figure}[h!]
	\begin{center}
		\subfloat[][30-days maturity. The mean is 0.448 with a standard deviation of 0.059]{\includegraphics[width=0.5\textwidth]{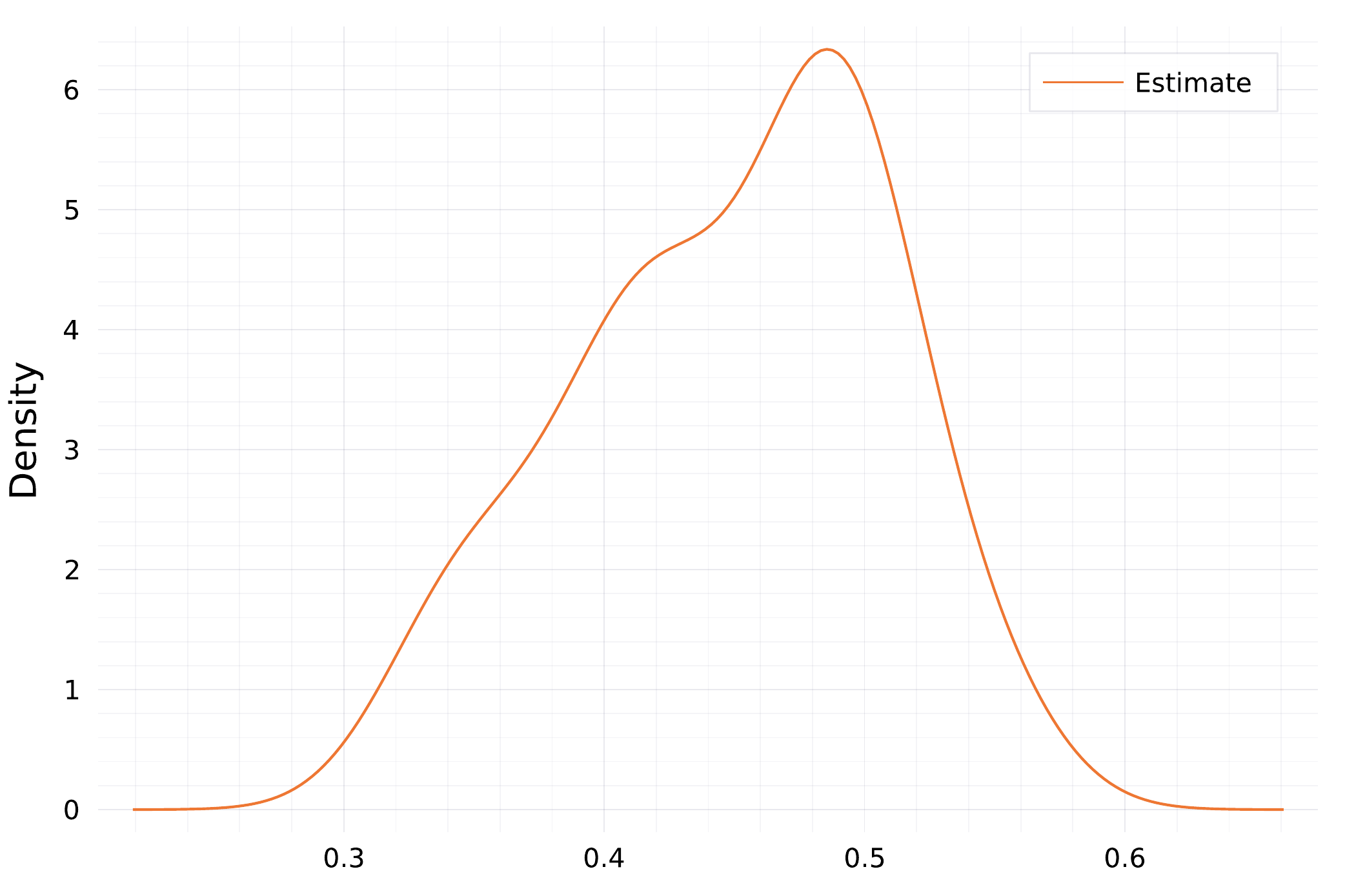}}
		\subfloat[][10-days maturity. The mean is 0.350 with a standard deviation of 0.049.]{\includegraphics[width=0.5\textwidth]{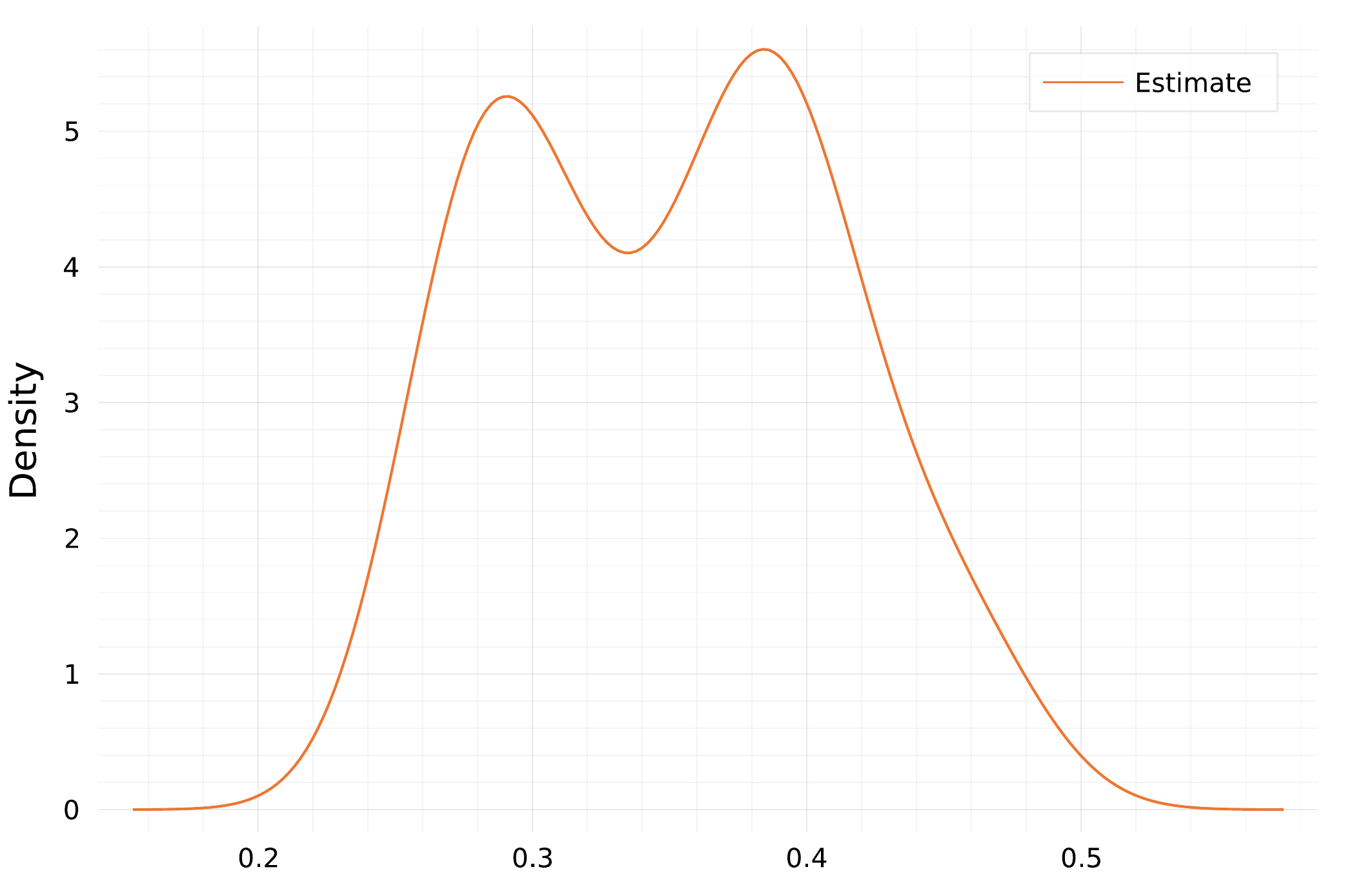}}
	\end{center}
	\caption{ MSFT  implied vol roughness index sliding estimate with $K=28$.}
\end{figure}

Roughness is more apparent in the 10-days implied volatility, with $H=0.350$. For the 60-days implied vol, we measure $H=0.416$ with a deviation of 0.06, so again not significantly different from the standard Brownian motion roughness. In comparison, \citet{livieri2018rough} measure a Hurst parameter of about 0.32 for options on the S\&P500 index with 30-days to maturity between January 5, 2006 and May 5, 2011.

The roughness of the implied volatility skew, also available in this data, is more pronounced.

\section{Measure of roughness of the VIX index}
Another interesting analysis is the roughness of the VIX index, which can be mapped to the price of a newly issued 30 days variance swap on the SPX index. It is slightly surprising that the various papers on rough volatility have not looked at it already. Using the quotes from Yahoo; with $K=51$, we measure $H=0.347$ with a standard deviation of 0.026. This is close to the 10-day implied volatility roughness measured previously.
\begin{figure}[h!]
	\begin{center}
		\subfloat[][from different starting dates.]{\includegraphics[width=0.5\textwidth]{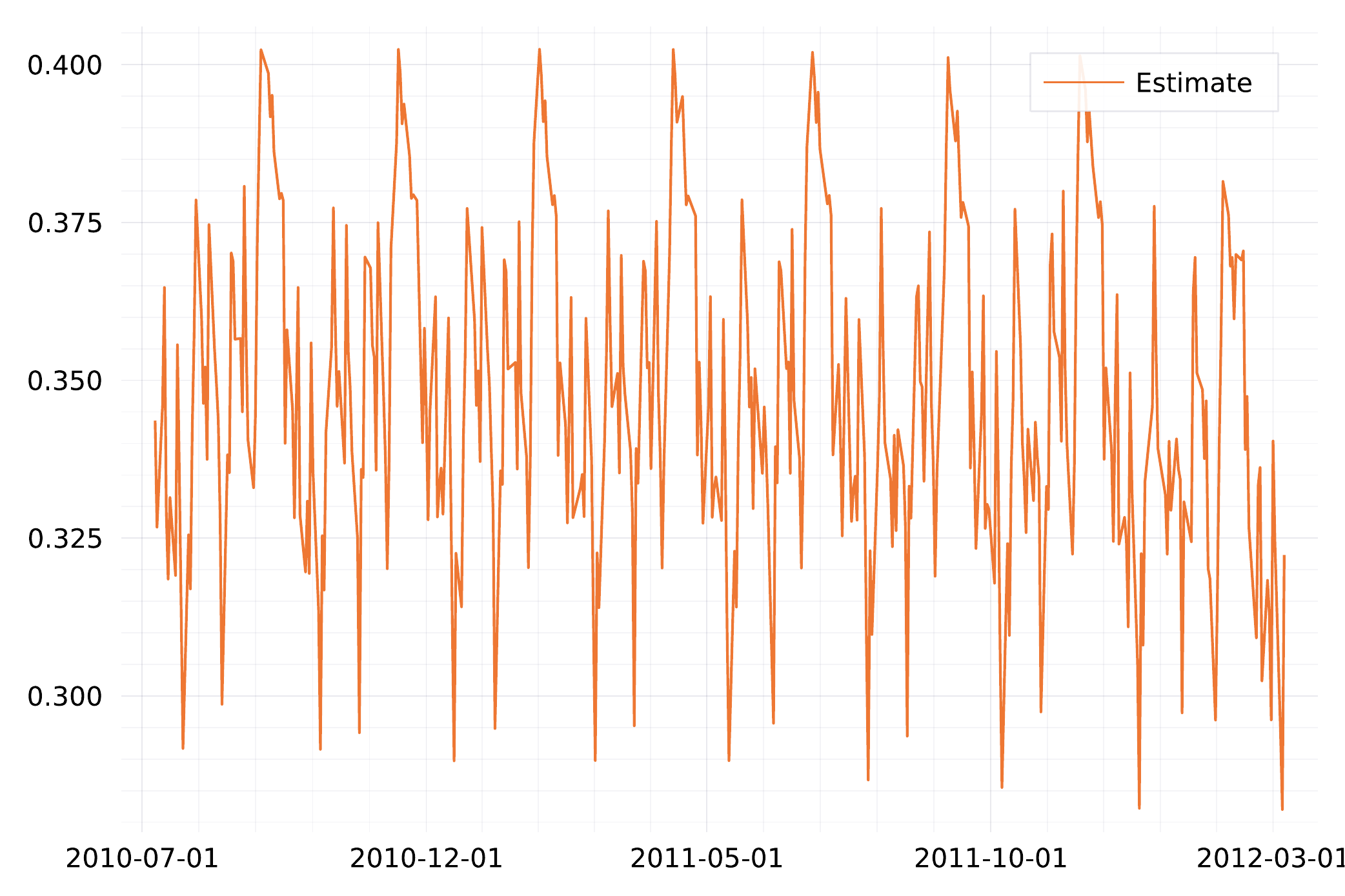}}
		\subfloat[][density.  The mean is 0.347 with a standard deviation of 0.026.]{\includegraphics[width=0.5\textwidth]{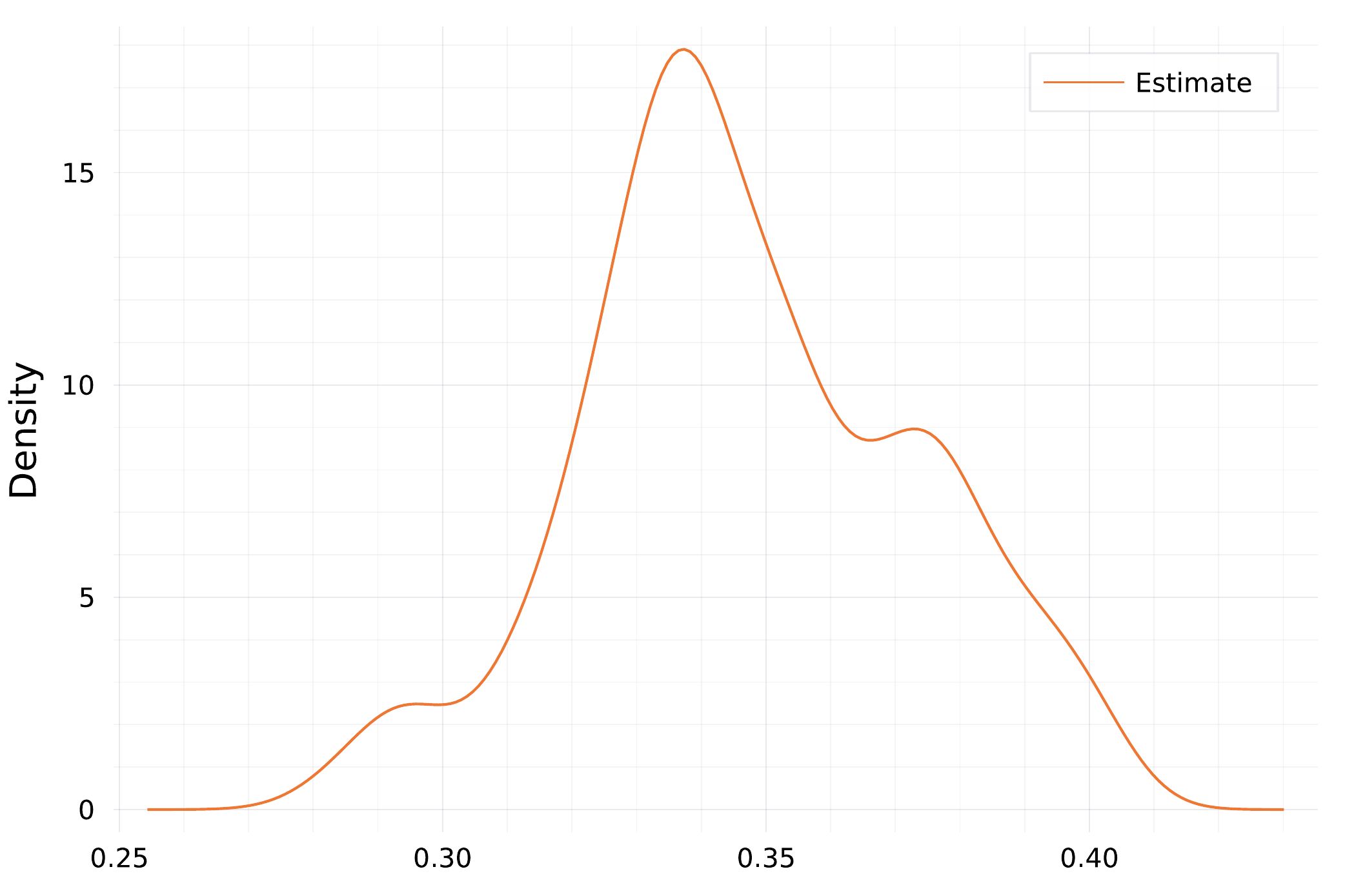}}
	\end{center}
	\caption{ VIX roughness index sliding estimate with $K=51$.}
\end{figure}

\section{Bias in the implied volatility proxy}
As in \citep{livieri2018rough}, we consider a rough volatility model where the asset price process $F$ follows
\begin{align}
	\frac{\diff F(t)}{F(t)} = \sigma v(t) \diff W_F(t)\,, &\quad \diff \ln v(t) = \eta \diff W_v^H(t)\,,
\end{align}
where $W_F$ is a Brownian motion, and $W_v^H$ is a fractional Brownian motion with Hurst index $H$, independent of $W_F$.

We discretize the system by the Euler scheme with a given time-step size $\Delta t$ (unless specified otherwise, we choose $\Delta t=0.001$) and compute the implied volatility at each day $t_i$ by inverting the option price obtained by simulating $M$ new paths from $t_i$ until $t=4$.  We then take the subset of values corresponding to a time-step of 0.004 (corresponding to one day in the BUS/252 daycount convention), thus allowing for multiple intermediate time-steps in the simulation of each path.

The implied volatility is computed by inverting the average at-the-money option price $C$ on the future paths:
\begin{equation}
	C = \frac{1}{M} \sum_{m=1}^M C_{\textsf{BS}}\left(1,1,\tau, \bar{w}) \right)\,,\label{eqn:call_price_mc}
\end{equation}
where $C_{\textsf{BS}}(F, K, \tau, w)$ is the Black-76 price of an option of strike $K$ and time to maturity $\tau$ on an asset of forward $F$ with total variance to maturity 
\begin{equation}
	\bar{w} = \int_0^\tau \sigma^2 v^2(s) \diff s\,.\end{equation}
There are several way to discretize the total variance. \citet{livieri2018rough} use a right rectangular rule:
\begin{equation}
	\bar{w}_R = \sum_{p=i+1}^{i+k} \sigma^2 v(t_p)^2 \Delta t_p\,,
\end{equation}
with $\Delta t_p = t_{p}-t_{p-1}$.
A trapezoidal rule may be more accurate for a given time-step size:
\begin{equation}
	\bar{w}_T = \sigma^2\left(  \frac{1}{2} v(t_i)^2  \Delta t_{i+1} + \frac{1}{2} v(t_{i+k})^2  \Delta t_{i+k} +  \sum_{p=i+1}^{i+k-1}  v(t_p)^2  \Delta t_p \right)\,.
\end{equation}
In order to calculate $v(t_p)$ we follow \cite{livieri2018rough}, and simulate the fractional Brownian motion for $t_j \in \left\{ t_{i+1},...,t_{i+k} \right\}$ using 
\begin{equation*}
	W_v^H(t_j) = \sum_{p=1}^i l_{jp}X_{00p} + \sum_{p=i+1}^{j} l_{jp}X_{mip}
\end{equation*}
where $l_{ij}$ are the elements of the lower triangular matrix in the Cholesky decomposition of the covariance matrix associated to $W_v^H$,  $X_{00p}$ are independent Gaussian random numbers representing the initial path, and $X_{mip}$ are independent\footnote{The independence between random numbers at different times $t_i$ is important to calculate the roughness of a single path (similar case as $M=1$), but is not necessary to compute the implied volatility roughness, as, at each time-step an expectation (represented by the averaging of $M$ paths with $M$ large enough) is calculated, which effectively "cancels" out the dependencies across different $t_i$ and thus only $N+Mk$ independent random numbers are necessary where $N$ is the number of time-steps in a full path. The initial path and the $M$ sub-path at each $t_i$ may then be further decoupled, allowing for the use of scrambled Sobol sequences for each.} Gaussian random numbers representing the $M$ paths starting at $t_i$. Then we have $v(t_p)= e^{\eta W_v^H(t_p)}$.

We justify our choice of time-step size, by measuring the roughness index of the implied volatility on multiple time-step sizes for a model with $\eta = \sigma = 0.5$ and $H=0.10$.  Table \ref{tbl:Hestimates20} shows that $\Delta t = 0.001$ provides sufficient accuracy when the trapezoidal rule is used to compute the total variance. Left or right rectangular rules require a much smaller time-step to provide the same accuracy and converge to the same estimate when $\Delta t \to 0$.

\begin{table}[h!]
	\centering{
		\caption{Estimate (standard error) of the roughness of the one-day implied volatilities, for several total variance discretizations in the rough exponential model with $\eta = \sigma = 0.5$, $H=0.1$ using 20 initial paths and $M=8192$ paths with antithetic variates, for each initial path. The standard error comes from the statistics on the initial paths. \label{tbl:Hestimates20}}
		\begin{tabular}{cccc}\toprule
			$\Delta t$ &  Trapezoidal & Right rectangular & Left rectangular \\	\midrule		
		0.001  & 0.258 (0.004) & 0.293 (0.004) & 0.225 (0.004)\\
		0.0004 & 0.255 (0.004) & 0.266 (0.004) & 0.238 (0.004)\\
		0.0002 & 0.258 (0.004) & 0.264 (0.004) & 0.257 (0.004)	\\
			\bottomrule
	\end{tabular}}
\end{table}

\begin{figure}[h!]
	\begin{center}
	\includegraphics[width=\textwidth]{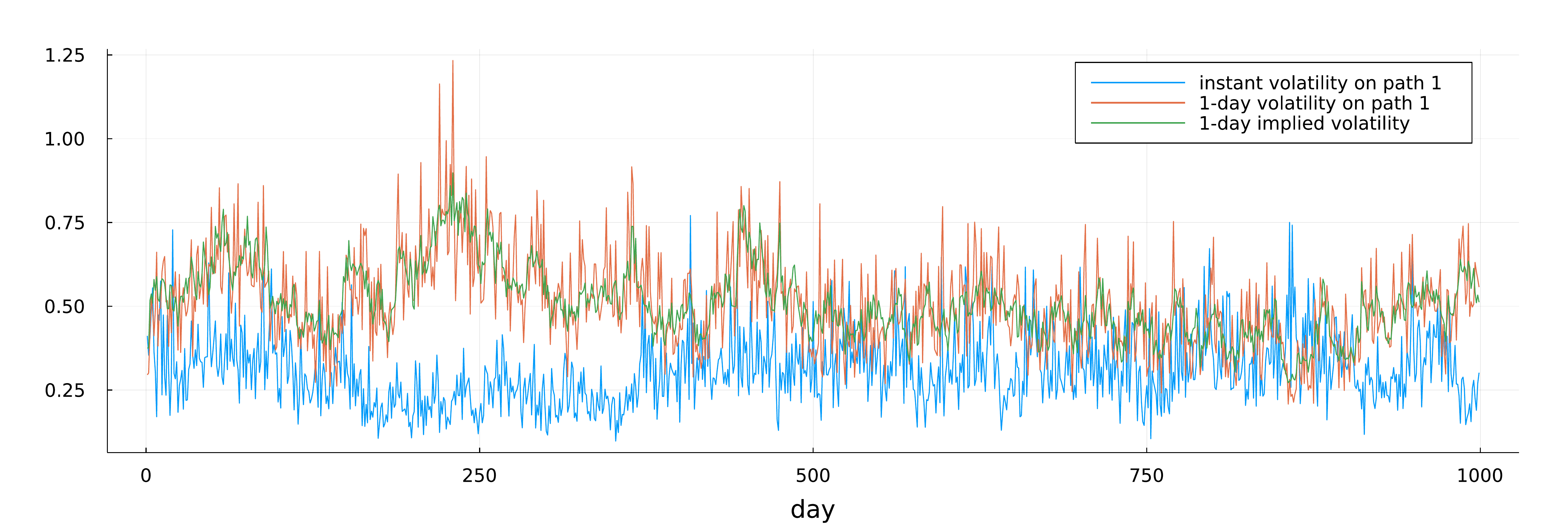}
	\end{center}
	\caption{Instant, integrated and 1-day implied volatility every day for the rough exponential model with  $\eta = \sigma = 0.5$ and $H=0.05$.\label{fig:vol_path_roughexp}}
\end{figure}

We consider the model with $\eta = \sigma = 0.5$ and $H=0.05$ and estimate the roughness index of the instantaneous volatility on one path, of the integrated one day volatility on one path, and of the one-day implied volatility, every day. Figure \ref{fig:vol_path_roughexp} suggests that the implied volatility is less rough than the integrated or instant volatilities.

\begin{table}[h!]
	\centering{
		\caption{Estimates of the roughness in the rough exponential model with $\eta = \sigma = 0.5$, using one initial path and $M=8192$ antithetic paths, for each initial path. The standard deviation comes from the sliding window estimate on 365 points. \label{tbl:Hestimates}}
	\begin{tabular}{cccc}\toprule
	maturity & volatility type & $H=0.05$  & $H=0.10$ \\
&&	estimate (std dev) & estimate (std dev)\\\midrule
	0 & Instantaneous  &0.079 (0.04) & 0.115 (0.04)\\\cmidrule(lr){2-4}
	1 business day &integrated, on path 1 & 0.102 (0.05) & 0.116 (0.06)\\
 &integrated, on average & 0.101  & 0.131 \\
 & implied & 0.218 (0.05) & 0.255 (0.05) \\\cmidrule(lr){2-4}
 	10 business days &integrated, on path 1 & 0.083 (0.04)& 0.099 (0.04)\\
 &integrated, on average & 0.095 & 0.108 \\
 & implied & 0.360 (0.04) & 0.373 (0.04)\\\cmidrule(lr){2-4}
 	20 business days  & implied & 0.392 (0.04) & 0.403 (0.03)\\
 
 \bottomrule
	\end{tabular}}
\end{table}

\begin{figure}[h!]
	\begin{center}
		\subfloat[][linear regressions for different $q$ values]{\includegraphics[width=0.5\textwidth]{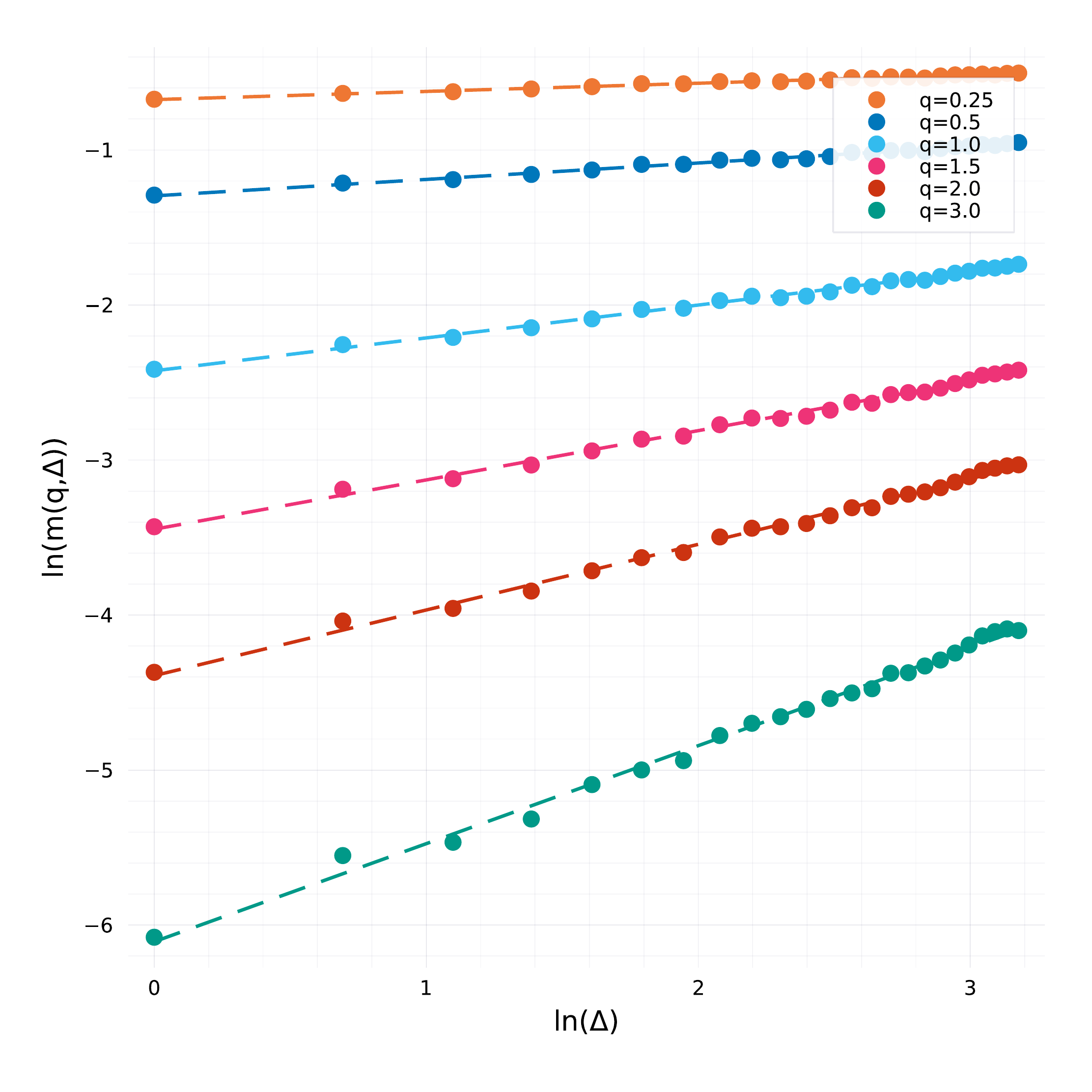}}
		\subfloat[][estimated $H=0.211$]{\includegraphics[width=0.5\textwidth]{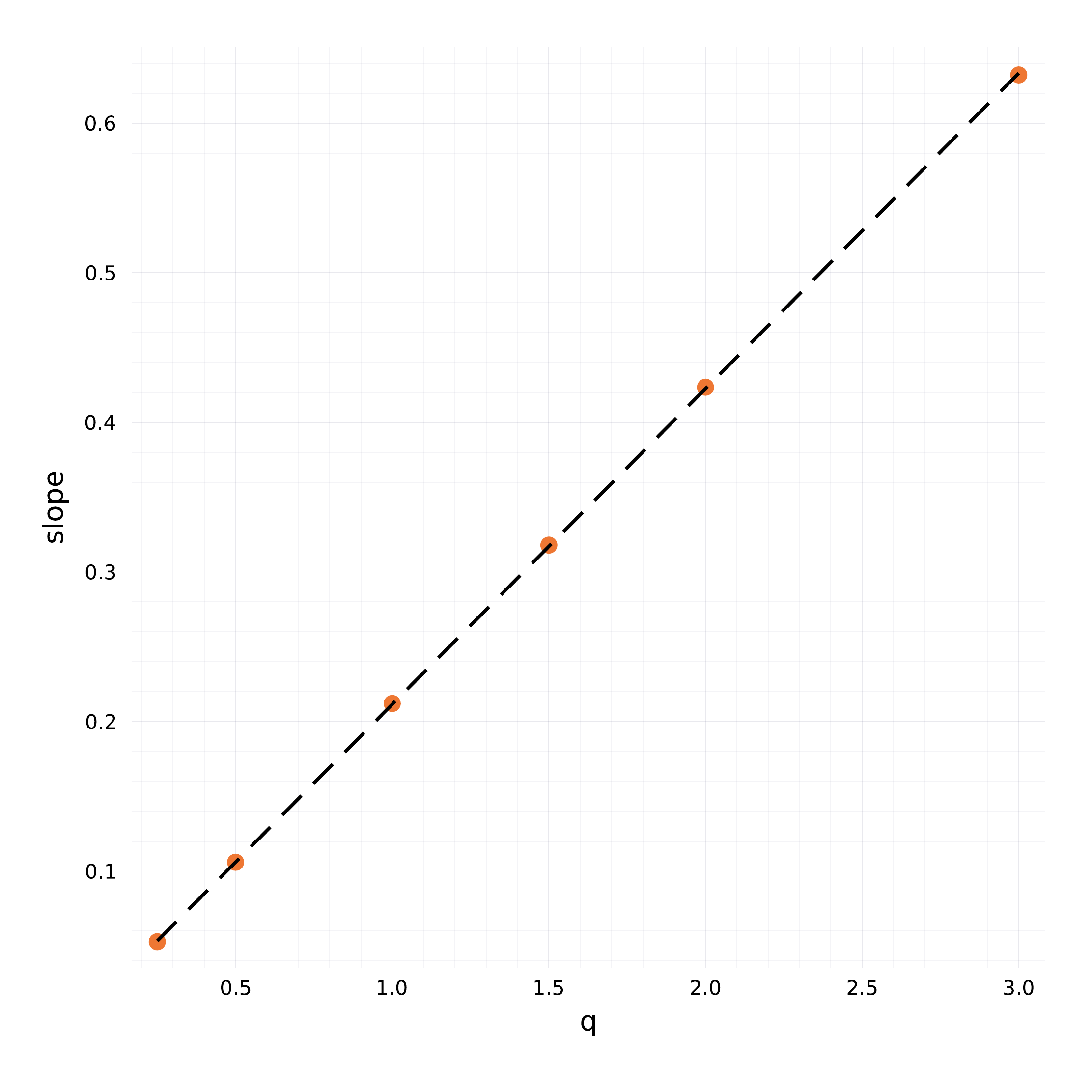}}
	\end{center}
	\caption{Estimate of $H$ using the linear regression method of \citet{gatheral2018volatility} on the 1-day implied volatilities with a model $H=0.05$. \label{fig:1day_ivol_roughexp_gatheral}}
\end{figure}

From Table \ref{tbl:Hestimates}, with $H=0.218$, the one-day implied volatility is found to be significantly less rough than the instantaneous volatility and significantly less rough than the estimate $H=0.06$ measured in \citep{livieri2018rough} for a similar setup. Our results are not impacted by the choice of measure (Equation \ref{eqn:H_pestimate}): a log regression as in \cite{gatheral2018volatility} leads\footnote{The method is relatively sensitive to the choice of $\max(\Delta)$ on which the regression is done. We choose here $\max(\Delta)=n/40  \approx 25$ where $n$ is the number of observations.} to a roughness $H=0.211$ (Figure \ref{fig:1day_ivol_roughexp_gatheral}), very close to the estimate via Equation (\ref{eqn:H_pestimate}).
The discrepancy with \citet{livieri2018rough} may be explained by the fact that we use intermediate time-steps in the simulation, while \citet{livieri2018rough} may not (and may use a single time-step per day).

The integrated variance on each path is of the same roughness as the instantaneous variance, regardless of the option maturity considered. The smoothing effect is introduced by the Monte-Carlo averaging over the paths via Equation (\ref{eqn:call_price_mc}), which corresponds to the expectation of the option price at a given day $t_i$.

The Hurst estimate of $0.35$ measured from the market implied volatilities suggests that the actual Hurst index of some corresponding underlying rough volatility process is smaller than $H=0.05$. This is consistent with the observations of \citet{fukasawa2019volatility}. There is big caveat: the uncertainty in the estimate. In our simulated rough exponential model with $H=0.05$, the minimum Hurst index  estimate of the 10 days implied volatilities is $H_{\min} = 0.33$ and the maximum is $H_{\max}=0.43$, over 10 initial paths of 4 years duration (a similar duration as our historical quotes analysis). Our procedure for estimating the Hurst index of some single historical path contains a large uncertainty (here around $\pm 0.05$). It will be more pronounced (in relative value) for low Hurst indices.
 
\begin{figure}[h!]
		\begin{center}
		\includegraphics[width=0.5\textwidth]{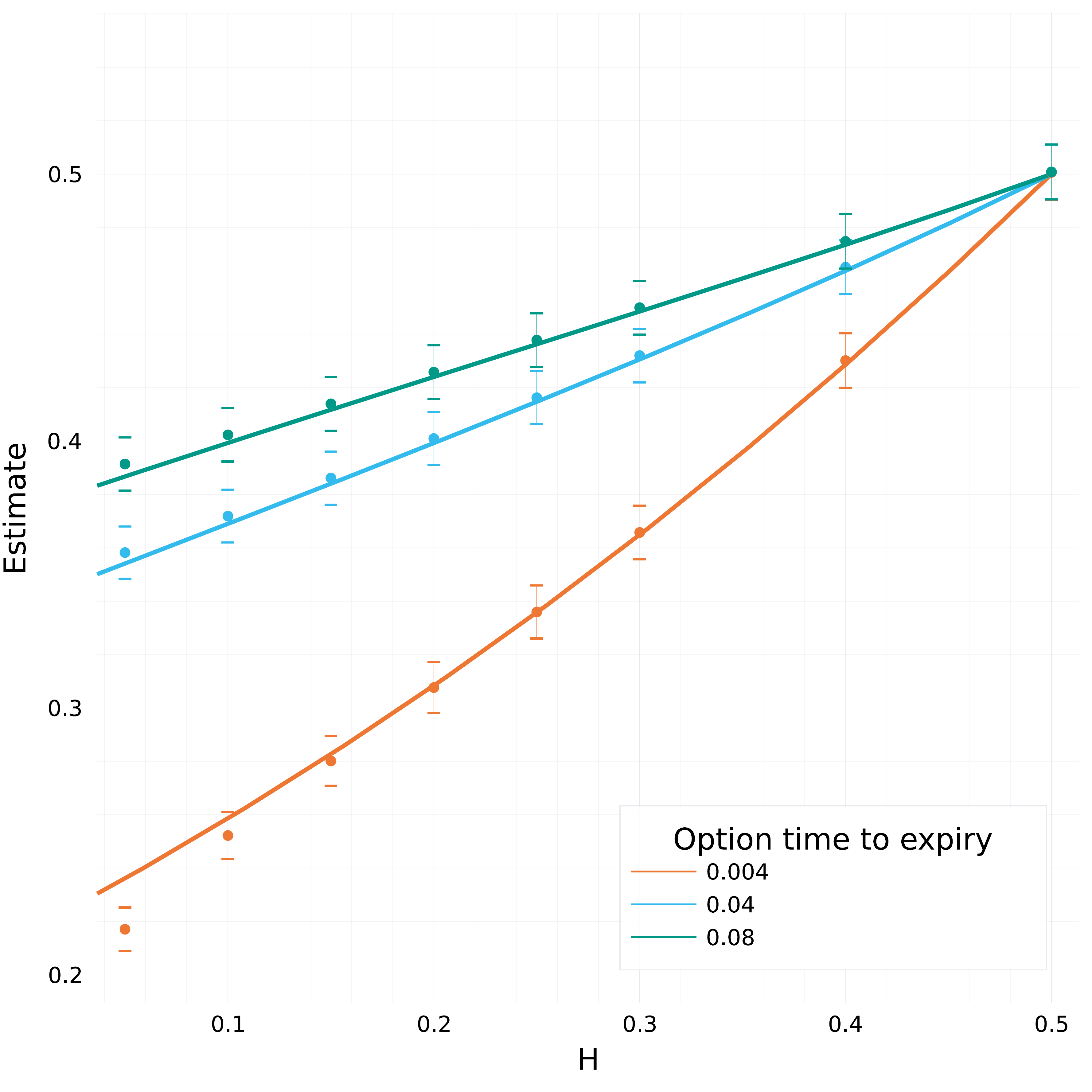}
		\end{center}
	\caption{Estimates of $H$ from the daily implied volatilities in the rough exponential model with  $\eta = \sigma = 0.5$ against the model $H$ for options with maturities of 1 day, 10 days and 20 days.\label{fig:HestimateH} Lines represent the theoretical value. Dots represent the quasi Monte-Carlo mean estimate over the initial paths, along with the 95\% confidence interval. The 32 initial paths are computed using a scrambled Sobol low discrepancy sequence.}
\end{figure}

It is interesting to notice that the relationship between the estimated $H$ of the implied volatility and of the true underlying model $H$ looks linear, regardless of the option time to expiry (Figure \ref{fig:HestimateH}). 
The slope of the line flattens with the option time to expiry, which implies a less precise estimate of the true underlying $H$ as the time to expiry increases. For options with an expiry of 20 business days, an error of 0.05 in the estimate implies an error of 0.10 of the underlying $H$. It appears then, that even for an expiry of 10 days, the error of the implied volatility proxy may be too large for practical purposes.
In \citep[Section 4.2]{livieri2018rough}, the theoretical relation is derived between the expected average volatility Hurst index and the model Hurst index, under slightly simplified assumptions on the model. The relation is not linear for a given time-to-expiry, and the non-linearity is more pronounced for small $H$ and small $T/K$ where the option expiry $T$ is measured in days. In Figure \ref{fig:HestimateH}, we plot the theoretical estimate of the expected volatility for a time-to-expiry $T$ in days
\begin{equation*}
	 \hat{H} = \frac{\ln(\hat{f} (T/K)) - \ln(\hat{f} (T))}{2\ln(K)} + H \,,
\end{equation*}
where $\hat{f}(\theta) = f_1(\theta) + f_2(\theta)$ with $f_1, f_2$ given in \citep[Section 4.2]{livieri2018rough}, and $K=25$ days. This corresponds to the slope of the linear regression of \citet{gatheral2018volatility} with $q=2$, using only the first and last observations. The theoretical estimate matches the results of the Monte-Carlo simulation well within the standard error.

The estimates presented in Table \ref{tbl:Hestimates} are nearly the same with with the rough Bergomi model. 

\section{Conclusion}
We reproduce the results of \cite{cont2022rough} and propose to use a sliding window estimate on top of their technique. The results suggests the following remark: while looking at the 5 minutes historical volatility may make sense to price intraday options,  it is not clear what is the importance of it for traditional derivatives pricing (with maturities longer than a week). Does the historical micro-structure really matter there?

We find the Hurst index $H$ of the implied volatility, and of the VIX index to be not as low as the one of the 5 minutes historical volatility. The estimated roughness $H=0.35$ is closer to the standard Brownian motion roughness. In fact, for the 30-day implied volatility (with daily quotes), the measured roughness was even closer to 0.5. 

This roughness does not necessarily correspond to the underlying stochastic volatility process roughness. A strong upward bias towards 0.5 is present in the at-the-money implied volatility proxy. This mostly confirms the findings of \citet{livieri2018rough}, although we find that even for a maturity of one day, the bias is large. For low $H$ values, the implied volatility proxy does not allow to guess the true Hurst index within a reasonable accuracy. Our measure of 0.35 for the Hurst index of the 10-day implied volatility only tells us that an underlying rough volatility process has $H<0.2$ with high confidence.

\funding{This research received no external funding.}
\conflictsofinterest{The authors declare no conflict of interest.}
\externalbibliography{yes}
\bibliography{rough_implied_volatility.bib}
\appendixtitles{no}
%TODO add appendix with code samples.
\end{document}